\documentclass[pdflatex,sn-mathphys-num]{sn-jnl}% 
\usepackage{graphicx}%
\usepackage{array}%
\usepackage{multirow}%
\usepackage{amsmath,amssymb,amsfonts}%
\usepackage{amsthm}%
\usepackage{mathrsfs}%
\usepackage[title]{appendix}%
\usepackage{xcolor}%
\usepackage{textcomp}%
\usepackage{manyfoot}%
\usepackage{booktabs}%
\usepackage{algorithm}%
\usepackage{algorithmicx}%
\usepackage{algpseudocode}%
\usepackage{listings}%
\usepackage{float} %
\usepackage{subcaption}%

\theoremstyle{thmstyleone}%
\theoremstyle{thmstyletwo}%

\theoremstyle{thmstylethree}%

\begin{document}

\newcommand{\mug}{$\frac{\mu g}{m^3}$}

\title[Article Title]{Ground-Level Near Real-Time Modeling for PM2.5 Pollution Prediction
}

\author*[1,2]{\fnm{Zachary R.} \sur{Fox}}\email{zfox1@its.jnj.com}

\author[1]{\fnm{Janet O.} \sur{Agbaje}}\email{\{agbajejo, maguiredd, loganjs, davismr, hansonha\}@ornl.gov}
\author[1]{\fnm{Dakotah} \sur{Maguire}}
\author[3]{\fnm{Javier E.} \sur{Santos}}\email{jesantos@lanl.gov}
\author[1]{\fnm{Jeremy} \sur{Logan}}
\author[1]{\fnm{Maggie} \sur{Davis}}
\author[4]{\fnm{Rima} \sur{Habre}}\email{habre@usc.edu}
\author[5]{\fnm{Jim} \sur{VanDerslice}}\email{jim.vanderslice@utah.edu}
\author[1]{\fnm{Heidi A.} \sur{Hanson}}

\affil*[1]{\orgdiv{Oak Ridge National Laboratory}, \orgaddress{\street{1 Bethel Valley Road}, \city{Oak Ridge}, \state{TN} \postcode{37980}}}

\affil[2]{\orgdiv{Johnson \& Johnson}, \city{New Brunswick}, \state{NJ} \postcode{08933}}

\affil[3]{\orgdiv{Los Alamos National Laboratory}, \orgaddress{\street{PO Box 1663}, \city{Los Alamos},  \state{NM} \postcode{87545}}}

\affil[4]{\orgdiv{University of Southern California}, \orgaddress{\street{1845 N Soto Street}, \city{Los Angeles}, \state{CA} \postcode{90089}}}

\affil[5]{\orgdiv{University of Utah}, \orgaddress{\street{201 Presidents Circle}, \city{Salt Lake City}, \state{UT} \postcode{84112}}}


\abstract{
Air pollution is a worldwide public health threat that can cause or exacerbate a myriad of illnesses, including respiratory disease, cardiovascular disease, and some cancers. However, epidemiological studies and public health decision-making are still stymied by the inability to assess impacts of pollution exposure in near real time.
To address this challenge, developing and deploying accurate digital twins of environmental pollutants will enable timely data-driven analytics---a crucial step in the modernization of health policy and decision-making.
Although other models both predict and analyze fine particulate matter exposure, they often rely on a multitude of modeled input data sources and other data streams that are not regularly updated.
Another challenge stems from current models relying on predefined grids.
In contrast, our deep-learning approach interpolates surface-level PM2.5 (particulate matter $\leq2.5$ \mug) concentrations between sparsely distributed US Environmental Protection Agency monitoring stations in a grid-free manner.
By incorporating additional, readily available datasets---including topographic, meteorological, and land-use data---we improve the model's ability to predict pollutant concentrations with high spatial and temporal resolution.
This enables simple model querying at any spatial location for rapid predictions without computing over the entire grid.
To ensure robustness, we randomize spatial sampling during the training process to enable our model to perform well in both dense- and sparse-monitored regions.
This model is particularly well suited for near real-time deployment because its lightweight architecture allows for fast updates in response to streaming data.
Moreover, the model's flexibility and scalability allow it to be adapted to various geographical contexts and scales, making it a practical and functional tool for delivering more accurate and timely air quality assessments.
Its capacity to rapidly evaluate multiple scenarios can be especially valuable for decision-making during a public health crisis.}


\maketitle

\section{Introduction}\label{sec:introduction}

Air pollution is a worldwide public health threat that exacerbates cardio-respiratory disease and increases acute risks of morbidity and mortality~\cite{de2022ambient, hamanaka2018particulate, requia2018global}. Unfortunately, epidemiological studies and public health decision-making are still stymied by a lack of accurate predictions of exposure (at high spatial resolution), support for disease forecasting, and timely interventions~\cite{wang2024us, vilcassim2023gaps}.
Accurate, real-time models of environmental pollutants can enable data-driven analytics and scenario simulations that are crucial to the modernization of health policy and decision-making.
However, challenges to accurate PM2.5 (particulate matter $\leq2.5$ \mug) prediction are numerous.
First, ground-level data are sparse and spatially distributed in a nonuniform way.
The gold standard data often come from devices managed by the US Environmental Protection Agency (EPA).
These devices measure PM2.5 concentrations by using reference methods~\cite{us2009national}, and the data are calibrated and undergo rigorous quality control.
Other consumer-grade sensor networks have been rapidly improving, including the PurpleAir network \cite{barkjohn2021development, ardon2019measurements}.
These sensors require more extensive calibration, although much of the recent effort has gone toward assimilating them with EPA sensor data within AirNow~\cite{AirNow}, which is the official EPA resource for real-time air quality and risk communication.
To provide sub-daily (hourly), real-time Air Quality Index values and risk communication, AirNow uses the NowCast algorithm. This approach uses weighted averages of 12 hourly PM2.5 measurements~\cite{NowCast}, and interpolation is implemented with inverse-distance weighting at a 5~km resolution~\cite{AirNowInterpolation}.
Although useful, PM2.5 concentrations are determined by complex processes that involve weather, varying pollution sources, and human dynamics that may not be captured by simple spatial interpolation and averaging-based approaches. 

Our method aims to address key needs in the spatial interpolation of ground-level PM2.5 prediction to meet demands across the health and environmental sectors for scalable and timely air quality prediction.
Our approach offers several key methodological advantages over previous methods for modeling PM2.5 concentrations: flexible and rapid spatial querying, effective integration of multimodal data streams, and adaptive handling of heterogeneous monitoring density.
These characteristics make the model particularly well suited for real-world air quality monitoring applications and exposure assessments, in which sampling locations are irregularly distributed and computational efficiency is paramount. 

We propose the use of the Sensevier~\cite{santos2023development} attention-based method to interpolate surface-level PM2.5 concentrations across the United States.
This method has a flexible querying system that does not rely on predefined geospatial grids.
This allows one to compute PM2.5 levels at locations of interest or across arbitrary paths without worrying about grid alignment issues.
It also avoids common interpolation artifacts that occur when moving between grid cells, thereby yielding smooth spatial PM2.5 distributions while still respecting the spatial heterogeneity of PM2.5 in regions where it is relevant.
Notably, our highly flexible method does not require computation over an entire domain.
In other words, the computational cost scales to the data density rather than to the entire geographic area (e.g., country, large geographic region).
Although our method is based primarily on EPA monitoring stations, it also integrates multimodal data streams for PM2.5 interpolation.
The model incorporates both temporally lagged measurements from these stations and colocalized meteorological conditions.
It also includes land-use characteristics, population density, and topographical information.

\begin{figure}
    \centering
    \includegraphics[width=0.95\linewidth]{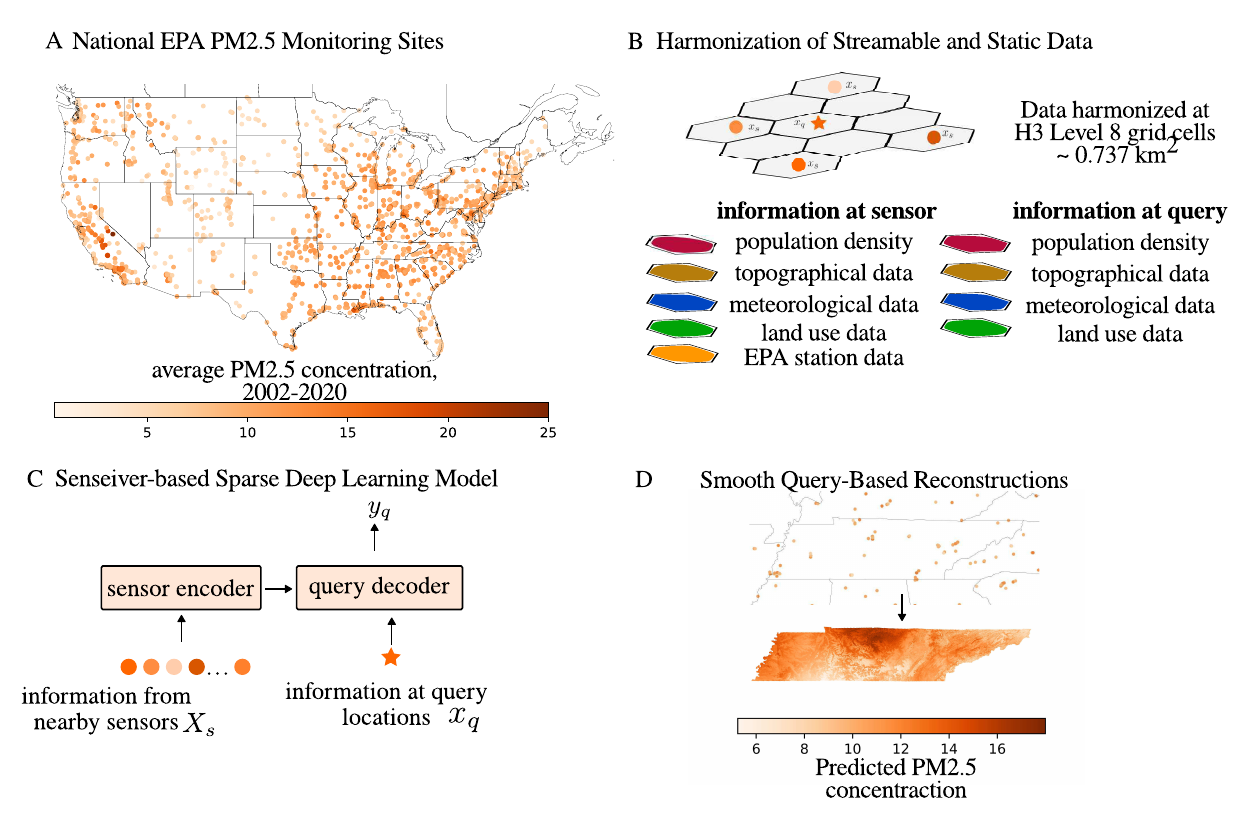}
    \caption{(a) Average daily PM2.5 concentration measured by 1,920 unique EPA monitoring stations in the contiguous United States (CONUS) during 2002--2020.
    (b) At each monitoring site, $x_s$, the static and streamable predictors are colocalized. PM2.5 levels are predicted at the query location, $x_q$, by using the subset of predictors available at the query location.
    (c) Sensor and query data are used as inputs to our deep learning architecture, which is based on the sparse-sensor attention-based Senseiver model.
    (d) After training, the model can perform interpolation at any resolution based on sparse sensor measurements. }
    \label{fig:Modeloverview}
\end{figure}

Finally, using a grid-free model provides more flexibility to account for the differences between urban and rural regions.
The model implicitly adapts its resolution to match local data density, thereby maintaining high accuracy in densely monitored urban areas while handling uncertainty in sparsely monitored rural regions (Fig.~\ref{fig:Modeloverview}).
Its capacity to evaluate multiple scenarios in real time based on recent changes in air quality and meteorology can be especially valuable for decision-making during a public health crisis.

\section{Previous Studies and Related Work}\label{sec:related_work}
Air pollution modeling has evolved significantly over the past several decades to encompass increasingly sophisticated spatiotemporal approaches.
The foundation for understanding PM2.5 health impacts was established through pioneering epidemiological work in the 1990s~\cite{schwartz1992philadelphia, dockery1993sixcities, schwartz1994children}, with seminal studies demonstrating robust associations between particulate air pollution and daily mortality and providing crucial evidence that formed the basis for modern air quality standards.
Early statistical approaches focused on integrating spatial interpolation with classical time series methods.
For example, Beckerman et al.~\cite{beckerman2013hybrid} developed a hybrid method that combines monthly PM2.5 observations with Bayesian maximum entropy techniques to demonstrate the value of incorporating multiple spatial scales in pollution modeling.
As computational capabilities advanced, machine learning approaches became increasingly prevalent.
In Thailand, Panneerselvam et al.~\cite{panneerselvam2023novel} conducted a comprehensive comparison of multiple machine learning models and found that Gaussian process regression achieved superior performance in capturing spatial and temporal patterns of air pollution.
This work highlighted the importance of model selection in different geographical contexts.
More recently, deep learning architectures specifically designed for environmental modeling have emerged.
Chen et al.~\cite{chen_novel_2023} used graph neural networks to model air pollution in Beijing, taking advantage of the natural network structure of monitoring stations.
Similarly, Xiao et al.~\cite{xiao2020improved} developed an improved long short-term memory network architecture and demonstrated an enhanced capability to capture temporal dependencies in pollution patterns.
The integration of multiple data sources has also become a key focus in the field.
Xiao et al.~\cite{xiao2018} demonstrated the power of ensemble models combined with satellite data, while Di et al.~\cite{di2019ensemble} integrated satellite, meteorological, land-use, and chemical transport model predictions in an ensemble learning system to produce daily 1~km PM2.5 estimates across the contiguous United States (CONUS), showing how diverse data streams can improve prediction accuracy.
This trend toward data-fusion approaches reflects the growing availability of complementary measurement systems and the need to leverage all available information sources.
Despite these advances, most existing approaches rely on grid-based methods that face trade-offs in compute and resolution.
Our work builds upon these foundations while addressing key limitations through a grid-free approach.

\section{Results}\label{sec:results}

\subsection{Daily Results for EPA PM2.5 Monitoring Stations from 2002 to 2020}
\begin{figure}
    \centering
    \includegraphics[width=.9\linewidth]{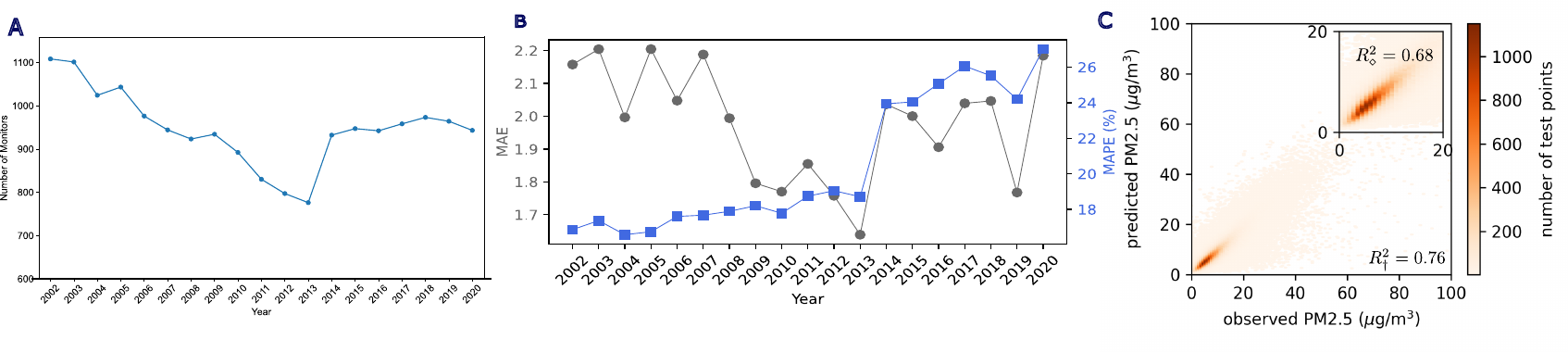}
    \caption{Overview of model results.
    (a)~PM2.5 monitoring stations in CONUS with data arranged by year.
    (b)~Yearly mean absolute error (MAE) (gray, left axis) and mean absolute percentage error (MAPE) (blue, right) for the model year over year.
    (c) Parity plot of the observed and predicted PM2.5 concentration across the CONUS from 2002 to 2020. $R^2_\dagger$ indicates the coefficient of determination for observed concentrations between 0 and 100 \mug. $R^2_\diamond$ indicates concentrations between 0 and 20 \mug.}
    \label{fig:all-results}
\end{figure}

With a study area of the CONUS and a date range of 2002--2020, the mean absolute error (MAE) on the test set is 1.98, which corresponds to a coefficient of determination, $R^2$, of 0.62.
Figure~\ref{fig:all-results}c shows the parity plot for the model performance.
The dense region of PM2.5 predictions occurs between 0 and 20 \mug (micrograms per cubic meter), and our model has an $R^2$ of 0.68 in that region (Fig.~\ref{fig:all-results}c, inset).
However, ignoring values greater than 100~\mug~results in an $R^2$ of 0.76 and an MAE of 1.7 \mug.
Next, we analyze the model's performance over time (Fig.~\ref{fig:all-results}b) by computing the MAE and mean absolute percentage error (MAPE) for the model's predictions on the test set for each year.
We find slightly decreasing performance over the study period, which is consistent with the modeling approaches of Di et al.~\cite{di2019ensemble}.
The MAPE shows a marked increase over time, whereas the average absolute error fluctuates between 2.2 and 1.6 over the study period. Generally, the average air pollution models decrease year-over-year, leading to a higher MAPE over time.
Notably, the number of unique monitors decreased from over 1,100 active monitors in 2002 to fewer than 1,000 in 2020 (Fig.~\ref{fig:all-results}a).
Although the number of monitors decreased, the number of observations per active monitor increased.
This change resulted in an increase in available data points---from 155,485 in 2002 to 256,319 in 2020. 

\begin{figure}
    \centering
    \includegraphics[width=0.8\linewidth]{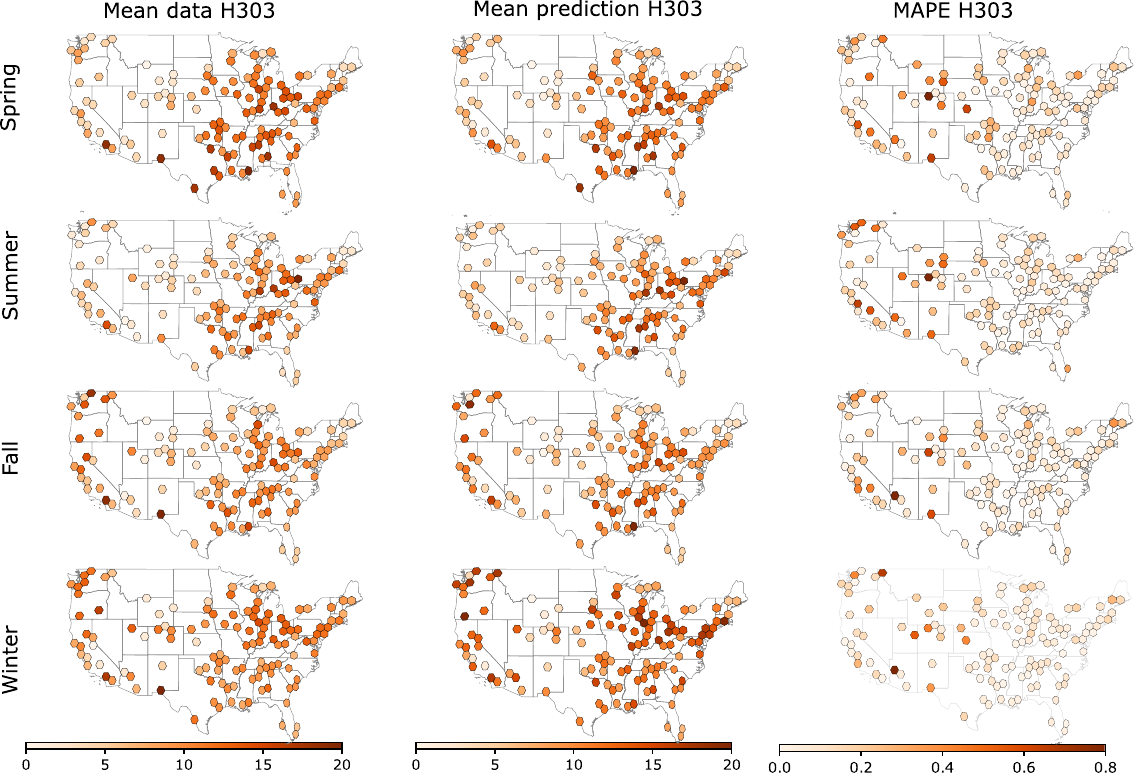}
   \caption{Model performance by season. Each monitor and associated prediction in the test dataset was averaged temporally over the study period and spatially aggregated at the H3 level 3 resolution (H303) in the left and center columns, respectively. The rightmost column shows the MAPE within each hex in \mug.}
    \label{fig:season-all-results}
\end{figure}

Our model also identifies seasonal variation (Fig.~\ref{fig:season-all-results}).
For example, in the summer, PM2.5 levels are elevated in most areas \cite{zhang2021using}.
And, as with other popular ground-level models of PM2.5 in the CONUS \cite{brokamp2022high,di2019ensemble}, our model is less accurate in the sparsely monitored western region.

\subsection{Uncertainty Quantification}
\begin{figure}
    \centering
    \includegraphics[width=0.6\linewidth]{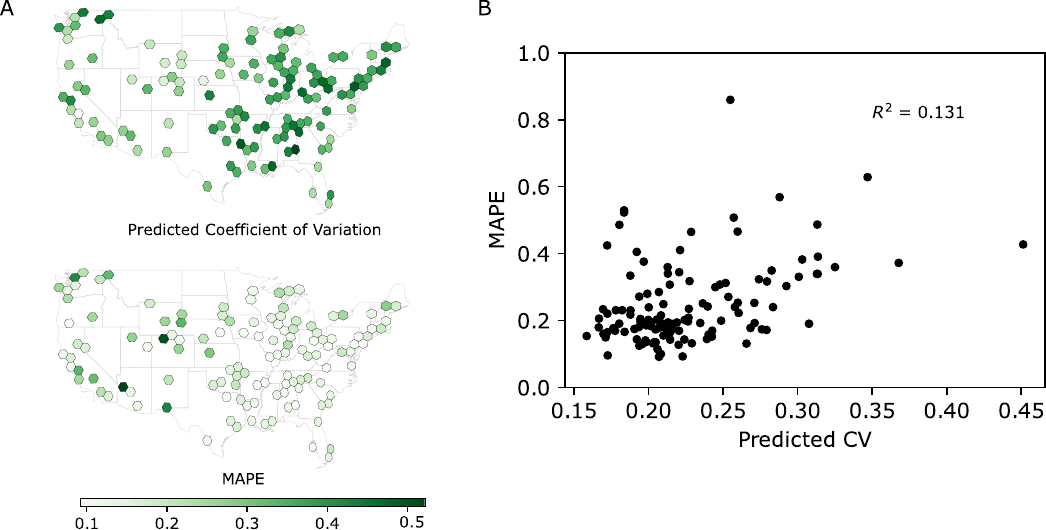}
   \caption{Uncertainty quantification for the model.
   (a) Predicted CV for the model (top) and the MAPE of the model over the entire study area at H303 resolution.
   (b) Predicted MAPE by CV spatially aggregated at the H303 level.}
    \label{fig:uq-results}
\end{figure}

The Senseiver architecture estimates uncertainty in the predictions by using different sets of monitors to generate predictions at inference time.
Figure~\ref{fig:uq-results} illustrates how well the model's uncertainty estimates align with its predictive performance.
The coefficient of variation (CV) is derived from 10 Monte Carlo simulations (see \nameref{sec:methods}) and provides a measure of relative uncertainty in model predictions.
These estimates are compared to the MAPE, which is calculated using ground-truth data.

Across the study area, the spatial patterns of predicted CV and MAPE show noticeable alignment, with many regions that exhibit elevated CV also showing higher prediction errors.
This suggests that the model’s internal uncertainty estimates reflect meaningful variations in predictive reliability across space.
Such spatial concordance is valuable for identifying areas where model outputs should be interpreted with greater caution.

The relationship between CV and MAPE is further quantified in the scatterplot, which shows a moderate positive Spearman correlation of 0.326.
This indicates a consistent monotonic relationship: higher uncertainty estimates are generally associated with greater prediction errors.
Although the linear fit explains only a small portion of the variance (\(R^2 = 0.131\)), the rank-based Spearman correlation reinforces the utility of CV as an uncertainty indicator, particularly when aggregated at H3 level-3 resolution (H303).

Together, these results suggest that the model's uncertainty estimates, although imperfect, are informative and can help identify regions with potentially reduced predictive accuracy.

\subsection{Feature ablation analysis}
\begin{figure}
    \centering
    \includegraphics[width=.7\linewidth]{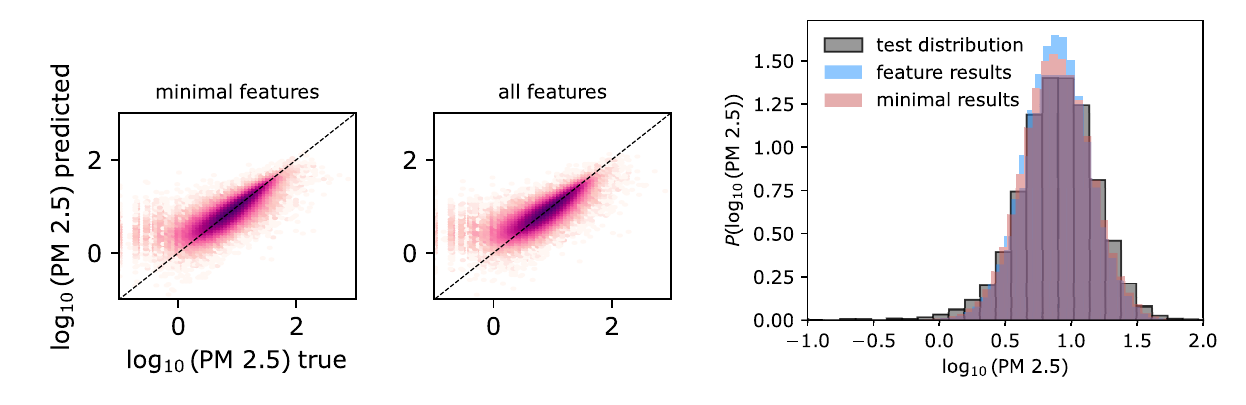}
    \caption{Left: Parity plots of all features and minimal feature sets.
    Right: Distribution of the predictions and true values.}
    \label{fig:minimal_vs_all}
\end{figure}

Next, we analyze why the model did not perform as well globally when using the full feature set vs. the minimal feature set.
We consider a \textit{minimal} feature set as one that includes PM2.5 concentration, land cover, time, latitude, longitude, and 15 daily lagged PM2.5 concentrations at the sensors along with latitude, longitude, and land cover at the query location. 
On the other hand, the \textit{full} feature set includes the minimal feature set (i.e., PM2.5 concentration, land cover, time, latitude, longitude, lagged PM2.5 concentration at the sensors) with the addition of air temperature, relative humidity, wind speed, precipitation, wind direction, day and night populations, elevation, and the same features without the PM2.5 features at the query locations.
 We first evaluate the performance of these two feature sets across the United States from 2002 to 2020 (Fig.~\ref{fig:minimal_vs_all}). 

\begin{figure}
    \centering
    \includegraphics[width=1\linewidth]{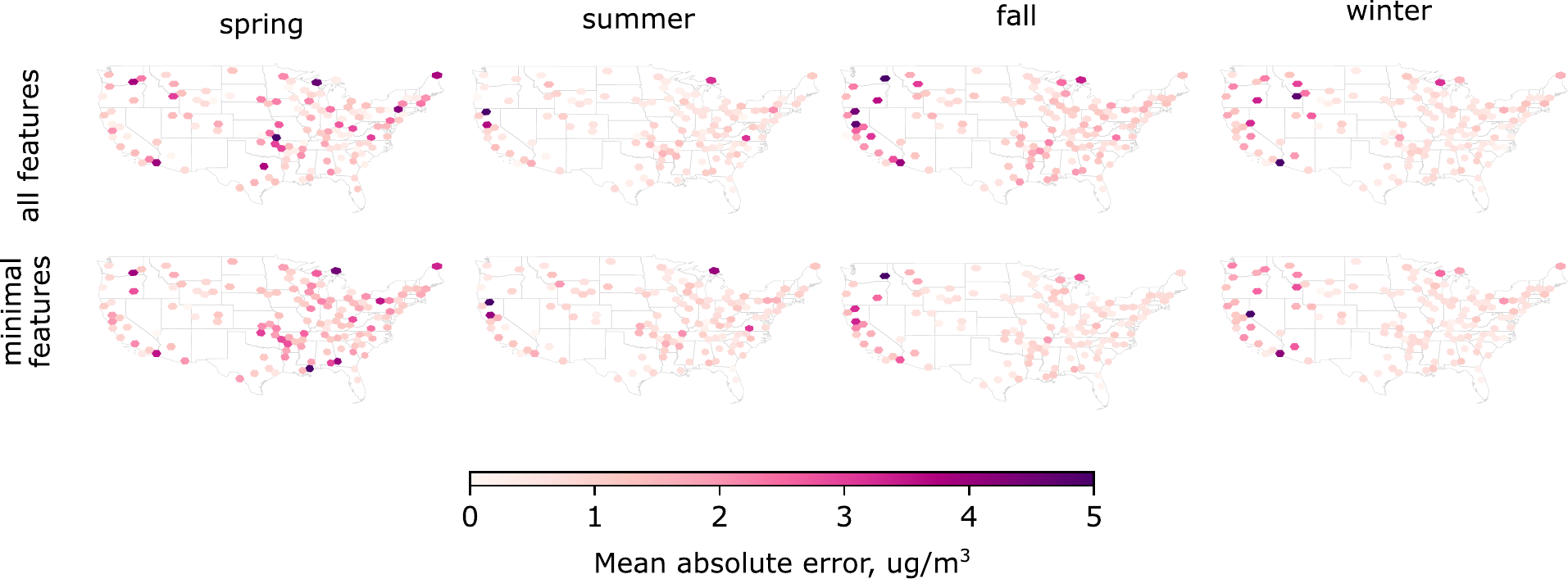}
    \caption{MAE of the predicted PM2.5 levels for each season. Top:~All features used in the model. Bottom:~Minimal features used in the model.}
    \label{fig:seasonal_minimal_vs_full}
\end{figure}

Next, we consider the effect of seasonality on the predictions for both the full and minimal feature sets (Fig.~\ref{fig:seasonal_minimal_vs_full}).
Although the same overall trends with respect to errors occur in both feature sets, the minimal feature set is not as accurate for the spring and performs better for the fall and winter, particularly in the western United States.
Notably, both feature sets are insufficient to capture trends for the spring. 

\subsection{Leave One State Out (LOSO)}
\begin{figure}[H]
    \centering
    \includegraphics[width=0.9\linewidth, trim=5 5 5 5, clip]{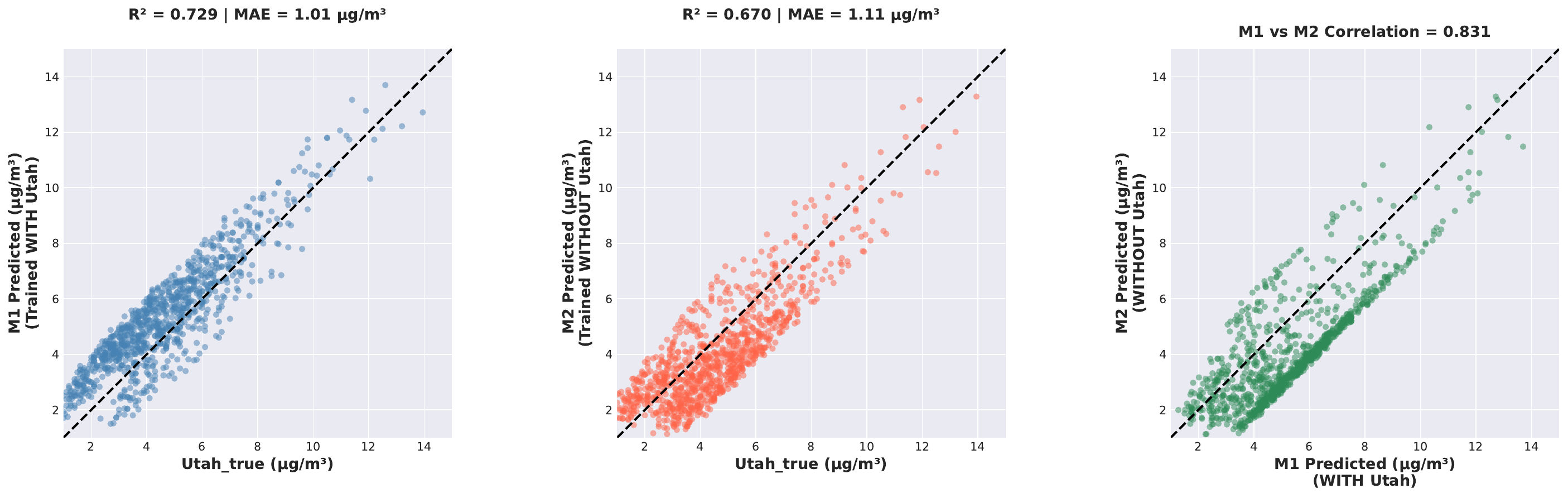}
    \caption{Model generalization across CONUS. LOSO with Utah as a case study. \textit{Left}: Model predictions on Utah data when trained on entire CONUS. \textit{Middle}: Model predictions on Utah data when trained with CONUS excluding Utah data points. \textit{Right}: Correlations between model predictions when trained with and without Utah data.}
    \label{fig:Utah_LOSO}
\end{figure}

To understand how the model generalizes across multiple geographic regions, we perform a spatial cross validation similar to ``leave one location out" (LOLO) \cite{brokamp2022high}, which we call ``leave one state out" (LOSO). Here, we trained a model only on data outside of a single state (Utah), and assessed how it performs on that state (Fig.~\ref{fig:Utah_LOSO}, middle). We chose Utah because the data are sparse and the terrain and meteorological conditions are complex. Utah exhibits dramatic elevation gradients, persistent winter temperature inversions, and highly localized precipitation regimes \cite{GilliesRamsey2012}, which together create challenging conditions for PM2.5 prediction. For the LOSO model evaluated on our Utah test set, we observe an $R^2$ of 0.670 and MAE of 1.11 $\mu g /m^3$ (Fig.~\ref{fig:Utah_LOSO}, middle). When we evaluate the model trained on all states on these same data points, we observe an $R^2$ of 0.729 and MAE of $1.01$ $\mu g /m^3$ (Fig.~\ref{fig:Utah_LOSO}, left). While there is slight degradation in performance, overall these predictions are highly correlated (Fig.~\ref{fig:Utah_LOSO}, right). We do not perform a full LOLO across multiple states due to computational costs associated with full cross validation in a deep learning model. 

\section{Methods}\label{sec:methods}
\subsection{Model}
The model used is based on the Senseiver architecture \cite{santos2023development}, which leverages cross-attention between sensor observations and query points to interpolate spatial fields without imposing a grid, thereby enabling flexible, data-adaptive predictions.
Although the original Senseiver was developed for general high-dimensional fluid dynamics problems,
 we adapted and streamlined the architecture specifically for ground-level PM2.5 prediction based on environmental sensor networks.

In our implementation, sensor inputs are composed of lagged PM2.5 concentrations, meteorological variables, land cover embeddings, elevation, and geospatial encodings, and query points include only static variables and location encodings.
Sensor and query features are concatenated early and passed into lightweight attention blocks that recycle parameters across layers to minimize model size.
During training, for each prediction, a target sensor is selected at random, and nearby sensors are dynamically sampled to ensure that the model learns robust interpolation patterns across a range of spatial densities.
We also extend the model with built-in uncertainty quantification at inference time by sampling multiple random subsets of sensors and aggregating predictions to provide spatially resolved uncertainty estimates without requiring separate ensemble models.
Loss functions are adapted to dynamically weight the contributions based on the number of sensors used in each batch.

\subsection{Uncertainty Quantification}
Let $\mathcal{X} \subset \mathbb{R}^d$ denote the spatial domain of interest.
Suppose we have $N$ sensors located at positions $\{x_1, x_2, \dots, x_N\} \subset \mathcal{X}$, each providing observations $y_i \in \mathbb{R}$ (e.g., temperature, pollution levels).
The goal is to predict the value $y^*$ at a new location, $x^* \in \mathcal{X}$, by using a neural network model, $f_\theta$, trained to perform interpolation based on a subset of sensor data.

Instead of utilizing all $N$ sensors, we randomly sample a subset, $S \subset \{1, 2, \dots, N\}$, according to a probability distribution, $P(S)$.
The model prediction at $x^*$, given subset $S$, is denoted by
\[
\hat{y}_S = f_\theta\left(x^* \mid \{(x_i, y_i)\}_{i \in S}\right).
\]

% \section{Monte Carlo Estimation of Predictive Uncertainty}

To quantify the uncertainty that arises from the choice of sensor subsets, we perform Monte Carlo sampling:

\begin{enumerate}
    \item Sample $M$ subsets $S^{(1)}, S^{(2)}, \dots, S^{(M)}$ independently from $P(S)$.
    \item For each subset $S^{(j)}$, compute the corresponding prediction:
    \[
    \hat{y}^{(j)} = f_\theta\left(x^* \mid \{(x_i, y_i)\}_{i \in S^{(j)}}\right).
    \]
\end{enumerate}

The ensemble of predictions, $\{\hat{y}^{(j)}\}_{j=1}^M$, enables us to estimate the predictive mean and variance at $x^*$:

\begin{align*}
\mu(x^*) &= \frac{1}{M} \sum_{j=1}^M \hat{y}^{(j)}, \\
\sigma^2(x^*) &= \frac{1}{M - 1} \sum_{j=1}^M \left(\hat{y}^{(j)} - \mu(x^*)\right)^2.
\end{align*}

Here, $\mu(x^*)$ serves as the aggregated prediction, while $\sigma^2(x^*)$ quantifies the epistemic uncertainty caused by sensor subset variability.

%\section{Interpretation and Theoretical Justification}

This Monte Carlo approach approximates the marginal predictive distribution:

\[
p(y^* \mid x^*) = \int p(y^* \mid x^*, S) \, p(S) \, dS,
\]

where $p(y^* \mid x^*, S)$ is the deterministic output of the neural network given subset $S$, and $p(S)$ is the sampling distribution over sensor subsets.
The variance $\sigma^2(x^*)$ thus captures the model's sensitivity to different sensor configurations to reflect the epistemic uncertainty.

\subsection{Training Procedure}
The training procedure is highlighted in Algorithm~\ref{alg:algorithm1}. 
This algorithm includes the sensor data that the model can use to predict the PM2.5 concentration at the query positions, $x_q$.
During training, the query positions must belong to the set of sensor locations because these locations have measurements associated with them.
However, at inference, any location in the CONUS can be queried.

To train this model, we follow a simple training procedure.
First, a query sensor is sampled.
The number of sensors is a hyperparameter that must be explored.
These sampled sensors will be used to predict the PM2.5 data at the query position.
They carry with them the meteorological features, population density, topographical data, and land-use information as well as spatial embeddings of the sensor locations and the PM2.5 measurements at the sensor locations.
The sampling is performed such that a given query is near the sensors.
How this sampling occurs must also be explored.
The query position carries with it the meteorological features and the spatial embeddings of the sensor locations.
The sensors and the query are used to predict the PM2.5 concentration at the query location.
Finally, gradients of the errors are used to update the model weights. 

 \begin{algorithm}
\caption{Training Procedure for Deep Neural Network}
\label{alg:algorithm1}
\begin{algorithmic}[1]
    \Require Training dataset $\mathcal{D}$, Encoder $E_{\theta}$, Decoder $D_{\phi}$, Number of sensors $N$, Learning rate $\eta$, Batch size $B$, Number of batches $N_B$, Number of epochs $N_E$
    \For{$e = 1$ to $N_E$}
        \For{$k = 1$ to $N_B$}
            \State Sample a batch of query points $\{q_1, q_2, \dots, q_B\} \sim \mathcal{D}$
            \State Sample $N$ sensor locations for each query point according to $\mathcal{N}(q_i, \sigma^2 I)$ without replacement
            \State Extract sensor features $X_i = \{x_{i1} \oplus x_{i2} \oplus \dots \oplus x_{iN}\}$ for each $q_i$
            \State Encode sensor features $Z_i = E_{\theta}(X_i)$ for each $q_i$
            \State Predict target values $\hat{y}_i = D_{\phi}(Z_i, x_{q_i})$ for each $q_i$
            \State Compute loss $\mathcal{L} = \frac{1}{B} 
            \sum_{i=1}^{B} \ell(\hat{y}_i, y_i)$
            \EndFor
        \State Update encoder and decoder weights:
        \State $\theta \leftarrow \theta - \eta \nabla_{\theta} \mathcal{L}$
        \State $\phi \leftarrow \phi - \eta \nabla_{\phi} \mathcal{L}$
        
    \EndFor
\end{algorithmic}
\end{algorithm}

\section{Data Sources and Preprocessing}\label{sec:data_sources}
\subsection{PM2.5 Data}

The PM2.5 concentration data that serve as our primary dataset originate from the EPA's Air Quality System, accessed via the Remote Sensing Information Gateway's Fused Air Quality Surface Using Downscaling files.\footnote{\url{https://www.epa.gov/hesc/rsig-related-downloadable-data-files\#output}}
This dataset comprises ground-based measurements from monitoring sites across the CONUS, with 6,832 unique observation days across 2002--2020.
It includes both 24-hour and 1-hour measurements that have been processed into 24-hour averages in units of \mug,~according rules governed by the National Ambient Air Quality Standards.
Each observation includes the following: 

\begin{itemize}
\item \textbf{Geographical coordinates:} Latitude, longitude, and site ID of 1,920 unique monitoring sites, which were reduced to 1,893 after filtering out stations outside the CONUS.
The sites were unevenly distributed across the CONUS, with a higher density of stations along the East Coast than the West Coast.
\item \textbf{Temporal data}: Date as daily timestamps.
\item \textbf{Concentration data}: Daily average PM2.5 concentration.
\end{itemize}

Not all monitoring sites were continuously operational throughout the study period, and many recorded data intermittently, often following a 3-day or 6-day sampling schedule.
Some locations had multiple colocated monitors that often recorded different concentrations on the same day and site and are distinguished by the EPA's Parameter Occurrence Code, which assigns unique identifiers to different instruments at the same location.
Following standard practice~\cite{brokamp2022high}, we aggregated colocated measurements by computing mean concentrations for identical spatial-temporal coordinates.
These data were integrated into Oak Ridge National Laboratory's (ORNL's) Centralized Health and Exposomic Resource, which serves as ground truth for model development.
 
\subsection{Elevation Data} 
To capture geographical influences on PM2.5 concentrations,
% aligning with our focus on features that change slowly or are static, 
we incorporated elevation values from the 2010 Global Multi-resolution Terrain Elevation Data (GMTED) \cite{Elevation} provided by the United States Geological Survey (USGS).
GMTED offers global digital elevation data at multiple resolutions, including 30-arc seconds ($\sim$1~km at the equator), 15-arc seconds ($\sim$500~m), and 7.5-arc seconds ($\sim$250~m) to ensure detailed terrain representations.

The elevation data at 30-arc seconds were re-gridded to level-8 H3 hexagons and extracted for each PM2.5 sensor location to provide a time-invariant geographical context that captures terrain-related influences on pollutant transport, dispersion, and deposition.

\subsection{Meteorological Data} 
Weather plays a critical role in PM2.5 transport, transformation, and removal processes.
To capture this influence, we integrated seven meteorological variables from the GridMet database, which provides daily gridded weather data at a 4~km resolution across the CONUS \cite{GridMet}.
These variables are spatiotemporally matched to PM2.5 location points and include daily minimum/maximum near-surface air temperature, daily minimum/maximum near-surface relative humidity, daily precipitation accumulation, and average wind speed and wind direction estimated at 10~m above the ground surface.

To ensure consistent spatial aggregation and modeling, we transformed the gridded GridMet data into H3 hexagons at resolution 8 by overlaying each raster pixel with its intersecting hexagons and computing area-weighted values.
For scalar variables (e.g., temperature, relative humidity, precipitation), we applied a conventional area-weighted averaging process based on the proportion of each GridMet cell falling within each H3 hexagon.
However, for wind data, we developed a vector-based approach to preserve directional integrity.
We decomposed wind measurements into U (east-west) and V (north-south) components, weighted these components by the proportion of each GridMet cell within each hexagon, and then reconstructed the resultant wind speed and direction from the aggregated components.
This approach avoids circular statistics problems that would arise from directly averaging the wind directions and ensures proper representation of wind vectors.
This hexification approach ensures spatial alignment across all variables and harmonizes input data for scalable, geospatially aware modeling.

The meteorological data were spatially indexed using Uber H3 resolution level-8 hexagons~\cite{uberh3} on the Andes compute cluster at ORNL.
This 704-node compute cluster enabled us to efficiently process multiple years of daily observations at high spatial resolution.

\subsection{Land Cover Data}
Provided by the USGS, the 2013 National Land Cover Database (NLCD) offers high-resolution (30~m) land cover classifications that capture critical features such as forests, urban areas, water bodies, and agricultural lands.
These classifications were aggregated to H3 level-8 hexagons ($\sim$0.74~km$^2$ resolution) to ensure spatial consistency with PM2.5 modeling grids.
Specific classes such as Dwarf Scrub, Sedge/Herbaceous, Lichens, and Moss were excluded because they primarily exist in Alaska~\cite{USGS_NLCD_Legend}.
The 2013 NLCD was selected because it represents a temporal midpoint in the study period and provides a balanced representation of land cover changes between older and newer periods.
Because land cover includes features (e.g., forests, urban areas, bodies of water) that typically change slowly over time, the 2013 dataset serves as a reliable and static representation of land surface characteristics that influence PM2.5 dispersion and deposition.

The land cover categories were first encoded as discrete integers and then mapped into a trainable embedding space with a dimension $N_l$ by using a learnable embedding layer within the model's encoder.
This transformation allows the model to capture complex relationships between land cover types by embedding them into a 12D space ($N_l = 12$), thereby facilitating effective integration with other spatial and temporal features.
These embeddings, which capture land surface characteristics, are concatenated with other features and processed through the model's attention mechanisms.
The spatial representation is further enhanced through Fourier encodings to capture multiscale periodic patterns, thereby enabling the model to learn relationships between land cover types and PM2.5 concentrations at different spatial scales.
We also ensure that the model prioritizes nearby land cover influences for local spatial sampling.

\subsection{LandScan Data}
The 2016 LandScan dataset was chosen to represent static features for the PM2.5 study spanning from 2002 to 2020.
This population density was developed by ORNL~\cite{landscan}.
LandScan provides high-resolution (1~km) estimates of daytime and nighttime population distributions and offers globally recognized, spatially detailed population metrics.
The 2016 dataset, derived from raw data at 3 arc-seconds ($\sim$90~m) resolution, was chosen because of its availability as the earliest year within the LandScan series (2016--2021) and its temporal overlap with the latter part of the PM2.5 study timeline (2002--2020).
For spatial consistency with the modeling framework, the population data were aggregated to H3 level-8 hexagons.
Population distribution trends are relatively stable over short periods, making the 2016 dataset a reliable representation of spatial exposure patterns during the study period.

\subsection{Lagged PM2.5 Variable}
To capture temporal dependencies in PM2.5 concentrations and very recent conditions, we engineered lagged features using a 15-day sliding window.
For each observation, we extracted historical PM2.5 values from the preceding 15 days to enable the model to learn from recent pollutant patterns.
Missing values in the temporal sequence were addressed through a two-stage interpolation approach.
First, we applied squared inverse distance weighting (IDW2) interpolation to identify neighboring monitoring sites with valid measurements on missing dates.
We searched for sensors at progressively increasing distances (5--50~km) and calculated interpolated values as distance-weighted averages of neighboring measurements (maximum of 32 sensors).
When IDW2 interpolation was insufficient because of sparse monitoring networks, we applied linear interpolation using the available time series data from the same site.
For locations with fewer than 15 available measurements in the window (particularly at the beginning of a time series), we padded the sequence with the current PM2.5 value to maintain consistent dimensionality.
Concentration values of less than 0.005 were dropped before logarithmic transformation.
This methodology ensured robust temporal feature representation---even in areas with sparse monitoring or irregular sampling---and significantly improved the model's ability to capture PM2.5 temporal dynamics.
% Hyperparameter Optimization: The number of lags were determined through hyperparameter optimization to balance model complexity with predictive performance.
% Feature Processing: The lagged values were truncated to retain only the first 15 values (nlags = 15) as specified in the configuration.

\subsection{Preprocessing Pipeline}
The preprocessing pipeline includes systematic handling of missing values and extreme value capping through upper/lower threshold scalers.
Data were merged on spatiotemporal keys (site ID and date) and partitioned into training (80$\%$), validation (10$\%$), and test (10$\%$) sets to ensure representative coverage across both spatial and temporal dimensions.
Normalized scaler parameters were saved for reproducibility.

\subsubsection{Data Normalization}
To ensure consistent scaling and improve model performance, we implemented a comprehensive normalization and standardization strategy tailored to each variable's characteristics.

\paragraph{Outlier Handling}
Outliers can disproportionately influence statistical measures such as mean and variance, leading to biased normalization and scaling.
To handle this, we introduced upper and lower thresholds that cap these values.

\paragraph{Variable-Specific Scaling}

\subparagraph{MinMax Scaler} A normalization was applied to the temporal and elevation features to ensure a fixed range [0, 1] for uniform scaling.
Negative elevation values were set to the lower threshold (0) for consistency:
\begin{equation}
scale = \frac{b - a}{x_{max} - x_{min}}
\label{eq:minmax_scale}
\end{equation}

\begin{equation}
x_{scaled} = (x - x_{min}) * scale + a,
\label{eq:minmax}
\end{equation}
where
$x$ is the original value,
$x_{min}$ is the minimum value in the dataset,
$x_{max}$ is the maximum value in the dataset, and
$a$ and $b$ are the feature range $[0,1]$, respectively.

\subparagraph{Standard Scaler} A ($Z$-score) normalization was applied to both meteorological and population data variables to standardize their mean and variance and avoid scale dominance in the model:
\begin{equation}
x_{scaled} = \frac{x - \mu}{\sigma},
\label{eq:standard}
\end{equation}
where $\mu$ is the mean and $\sigma$ is the standard deviation of the input data.

\subparagraph{Logarithmic Scaler} After establishing a minimum threshold of 0.001 to handle skewed distributions, a log-transform (base 10) was applied to the PM2.5 concentration and the lagged PM2.5 values:
\begin{equation}
x_{scaled} = \frac{\log(x)}{\log(b)},
\label{eq:log_scale}
\end{equation}
where $b$ is the base.

\subparagraph{LatLon Scaler} This scaler was used to convert geospatial coordinates (latitude and longitude) to normalized Cartesian coordinates and preserve spatial relationships.

\subparagraph{Spatial Encoding Pipeline} A two-stage process transforms geographical coordinates.
First, latitude ($\phi$) and longitude ($\lambda$) are converted to normalized Cartesian coordinates on a unit sphere:
\begin{equation}
\begin{aligned}
x &= \cos(\phi) \cos(\lambda) \\
y &= \cos(\phi) \sin(\lambda) \\
z &= \sin(\phi)
\end{aligned}
\label{eq:latlon_cartesian}
\end{equation}

\begin{equation}
\begin{aligned}
x_{norm} &= \frac{x}{\sqrt{x^2 + y^2 + z^2}} \\
y_{norm} &= \frac{y}{\sqrt{x^2 + y^2 + z^2}}.
\end{aligned}
\label{eq:latlon_normalize}
\end{equation}

These normalized coordinates are then projected onto a Fourier basis using sine-cosine positional encoding:
\begin{equation}
\begin{aligned}
a_i &= \sin(2\pi n_i x/P), \quad i = 1,...,N_f \\
a_{i+N_f} &= \cos(2\pi n_i x/P), \quad i = 1,...,N_f
\end{aligned}
\label{eq:fourier_encoding}
\end{equation}

\begin{equation}
x_{encoded} = [a_1, ..., a_{N_f}, \;\;\; a_{N_f+1}, ..., a_{2N_f}],
\label{eq:full_encoding}
\end{equation}
where $n_i$ represents the frequency bands from 1 to $N_f$, $P$ is the encoding period, and the final encoding concatenates sine components ($a_1$ to $a_{N_f}$) with cosine components ($a_{N_f+1}$ to $a_{2N_f}$) to create a rich multiscale spatial representation.

 %\begin{figure}[htp] 
    % \begin{center}
    % \includegraphics[width=6cm]{Picture2.png}
     %\caption{An n-dimensional point in space x is decomposed into a set of spectral features.}
     %\label{multiscale_fig}
    % \end{center}
% \end{figure}

\subparagraph{Categorical Scaler} Land cover classes were categorically encoded to reflect their discrete nature.

% The preprocessing pipeline includes systematic handling of missing values and extreme value capping through upper/lower threshold scalers. Data were merged on spatiotemporal keys (site ID and date) and partitioned into training (80$\%$), validation (10$\%$), and test (10$\%$) sets, ensuring representative coverage across both spatial and temporal dimensions. Normalized scaler parameters were saved for reproducibility. 

\section{Discussion}\label{sec:discussion}

Accurate spatial interpolation of PM2.5 concentrations is a critical capability for protecting public health.
Our study introduces a grid-free modeling approach that flexibly integrates multimodal data and adapts to heterogeneous sensor densities across space and time.
The proposed method demonstrates strong performance under a wide range of conditions and is well suited for real-world exposure assessments.

To meet these needs, we include relatively few model features.
Although aerosol optical depth (AOD) from satellite observations is commonly used as a proxy for surface-level PM2.5 after many adjustments, we chose to omit it from our feature set.
This decision was based on several considerations.
First, although AOD provides broad spatial coverage, it represents a column-integrated measure of aerosol concentration and is often poorly correlated with ground-level PM2.5, particularly in regions with complex meteorology or vertical aerosol stratification~\cite{zhai2021relating}.
Second, AOD availability is limited under cloudy conditions and varies seasonally, introducing non-random missingness that can bias learning.
Third, including AOD as a feature can inadvertently confound the model by encouraging it to learn spurious associations between satellite-derived signals and surface pollution levels, especially in data-sparse regions where direct ground-truth supervision is limited.
By excluding AOD, we focus on ground-based and contextual features that more directly reflect local pollution dynamics.
This choice aligns with our emphasis on interpretability, generalizability, and robustness in a grid-free predictive setting.

Despite a stable MAE over time, the MAPE (Fig.~\ref{fig:all-results}) increases, coinciding with a decrease in pollution levels across the United States (i.e., the MAE is constant, but the percentage of error increases as PM2.5 declines).
However, this is confounded by an increase in the number of measurements available in recent years, and we do not see model performance correlating with the amount of available information at a given spatiotemporal location.

Our model produces robust predictions across seasons and captures elevated PM2.5 concentrations in the summer months consistent with known trends~\cite{zhang2021using}.
However, performance degrades in the western United States, especially during spring.
This may be due to the sparse sensor network in these regions and the difficulty of capturing episodic events such as wildfires and dust storms, which can disproportionately impact spring air quality.
These patterns mirror performance characteristics of other high-resolution PM2.5 models~\cite{brokamp2022high, di2019ensemble} and suggest that sparsely monitored areas remain a challenge for all approaches.

We explored the effect of feature richness on model performance through an ablation study that compared \textit{full} and \textit{minimal} feature sets.
Interestingly, the model with the minimal feature set (lagged PM2.5 concentrations, land cover, location, and time) performed comparably to the model with the full set in aggregate, and---in some contexts---it even outperformed the full feature set.
Specifically, in the fall and winter, the model with the minimal feature set yielded lower error in western regions, whereas the full feature set underperformed in those same areas.

This finding underscores the importance of parsimony in model design.
Adding more features does not necessarily lead to better generalization and may introduce noise or exacerbate overfitting, particularly in regions with limited training data.
Moreover, these results emphasize that lagged local PM2.5 measurements, land use, and simple geospatial features contain significant predictive power when combined through an appropriate deep learning architecture.

A key advantage of our model is its inherent capability to provide uncertainty estimates via variable sensor combinations at inference time.
This allows for region-specific uncertainty quantification without requiring an explicit ensemble or Bayesian approximation.
Such predictive uncertainty is essential for risk-aware decision-making and downstream health impact modeling.

\begin{figure}[h]
    \centering
    \includegraphics[width=0.7\linewidth]{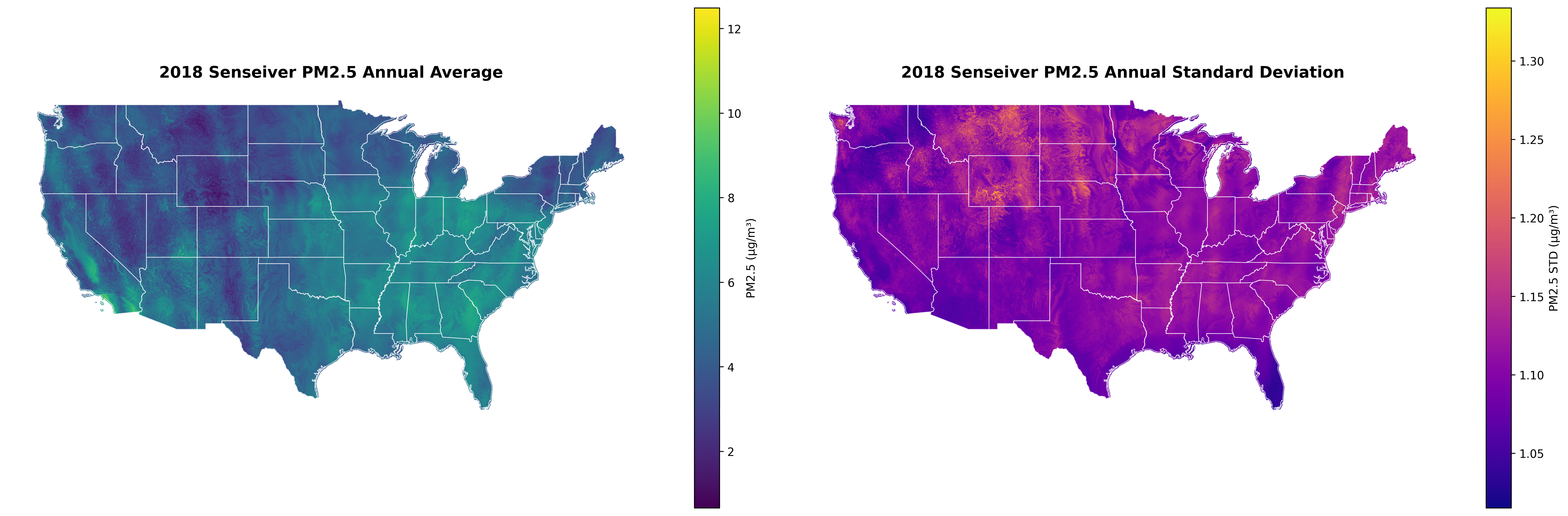}
    \caption{High-resolution Senseiver prediction of average annual PM2.5 concentrations for 2018 across CONUS (H308).}
    \label{fig:conus_pred}
\end{figure}

Our method represents a step forward in air pollution modeling, particularly in its ability to make fine-grained, data-adaptive predictions over irregular spatial domains.
The grid-free formulation allows for querying of pollution levels at arbitrary spatial resolutions and locations of interest, such as schools, hospitals, or residential zones.
The model produces continuous smooth prediction surfaces throughout the CONUS (Fig.~\ref{fig:conus_pred}), eliminating the discontinuities that can arise from traditional grid-based interpolation methods.
This makes the model especially attractive for health impact assessments and responsive public health policies.

However, limitations remain.
The performance of the model in data-sparse regions and during complex meteorological events (e.g., wildfire seasons, Santa Ana wind events) still lags behind more data-dense contexts.
Future work could focus on incorporating remote sensing data (e.g., satellite-derived AOD) or on developing hybrid models that fuse physical dispersion models with learned representations.
Additionally, more explicit modeling of temporal dependencies (e.g., through recurrent or diffusion-based dynamics) may further improve long-term predictive fidelity.

In conclusion, our results demonstrate the feasibility and utility of a grid-free, sensor-adaptive approach for PM2.5 interpolation.
Crucially, these models cannot rely on lagging computational features that come from data streams such as satellite imagery and slow-running computational models, which compromise our model's performance.
However, with continued refinement and expansion, such models can serve as powerful tools for real-time environmental monitoring and public health analytics.

\section{Model Validation}\label{sec:model_validation}
\subsection{IMPROVE PM2.5 Dataset}
The Interagency Monitoring of Protected Visual Environments (IMPROVE) network is a long-term air quality monitoring program established in 1985 by the EPA in collaboration with federal land management agencies and researchers at the University of California, Davis.
Its primary purpose is to monitor visibility conditions in national parks and wilderness areas across the United States to identify sources of visibility impairment and document long-term trends in atmospheric particulate matter.

\begin{figure}
    \centering
    \includegraphics[width=0.65\linewidth]{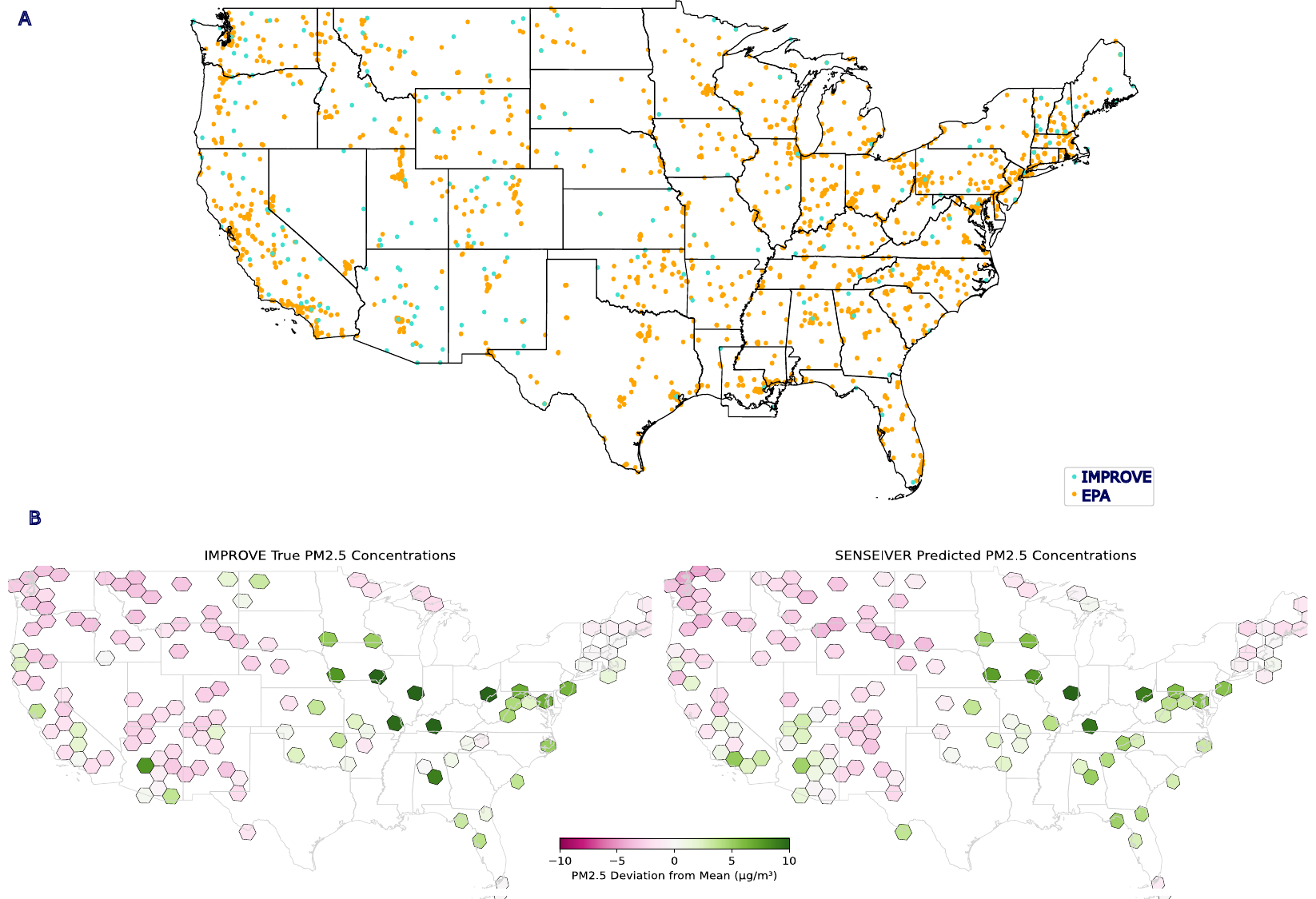}
    \caption{(a)~IMPROVE vs. EPA monitoring network location.
    (b)~IMPROVE vs. Senseiver PM2.5 concentrations.}
    \label{fig:improve_epa}
\end{figure}
The network comprises over 150 monitoring sites strategically located in Class I areas, including national parks, wilderness areas, wildlife refuges, and tribal lands that receive special levels of air quality protection for visibility under the Clean Air Act~\cite{EPA_PSD_ClassI}.
IMPROVE stations collect PM2.5 and PM10 samples via automated aerosol samplers that operate on a 1-in-3 day schedule, with each 24-hour sample providing detailed chemical composition data for particulate matter.
In contrast to the EPA's State and Local Air Monitoring Stations network, which focuses primarily on population-dense urban and suburban areas for regulatory compliance monitoring, IMPROVE sites are strategically positioned in remote, pristine environments with minimal local pollution sources.
This complementary spatial distribution makes IMPROVE data particularly valuable for validating air quality models in diverse geographical and environmental contexts.
Our analysis reveals that the IMPROVE and EPA monitoring networks have minimal spatial overlap, and IMPROVE captures air quality conditions in areas typically underrepresented by conventional regulatory monitoring (Fig.~\ref{fig:improve_epa}a).
We used IMPROVE PM2.5 observations collected from 2002 to 2020 ($\sim$122 sampling days per year) as an independent validation dataset, allowing us to assess model performance across both urban environments (EPA sites) and remote background locations (IMPROVE sites).
Figure~\ref{fig:improve_epa}b shows the comparison between observed IMPROVE concentrations and our model predictions with a \textit{full} feature set for February 16, 2014.

\subsection{Wildfire Phenomenological Case Study: Cameron Peak Fire}
\begin{figure}[H] 
    \centering
    \setlength{\belowcaptionskip}{-2pt}
    \begin{subfigure}[b]{0.30\textwidth}
        \centering
        \caption{}    
        \includegraphics[width=\textwidth, trim=0 20 0 5, clip]{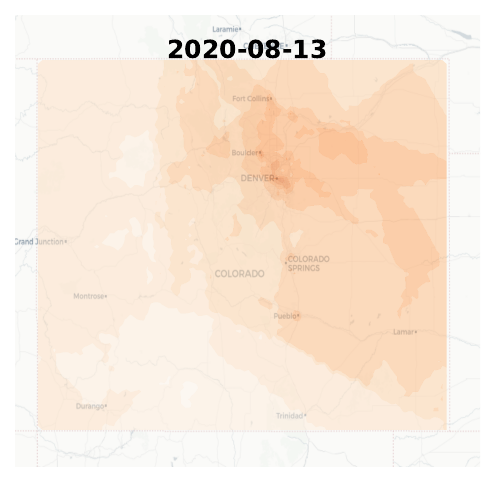}
        \label{fig:cameron_a}
    \end{subfigure}
    \hfill
    \begin{subfigure}[b]{0.30\textwidth}
        \centering
        \caption{}
        \includegraphics[width=\textwidth, trim=0 20 0 5, clip]{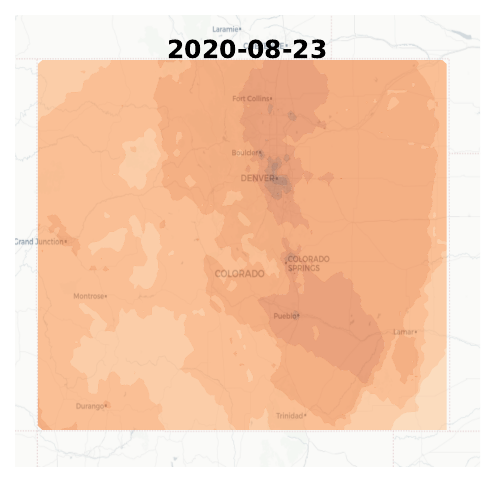}
        \label{fig:cameron_b}
    \end{subfigure}
    \hfill
    \begin{subfigure}[b]{0.30\textwidth}
        \centering
        \caption{}
        \includegraphics[width=\textwidth, trim=0 20 0 5, clip]{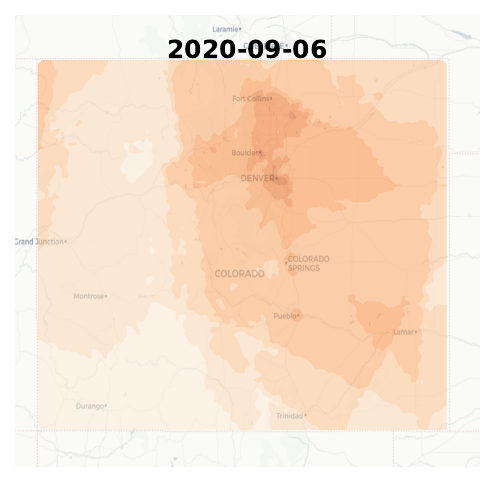}
        \label{fig:cameron_c}
    \end{subfigure}
     \vspace{-0.5em}
    \begin{subfigure}[b]{0.30\textwidth}
        \centering
        \caption{}
        \includegraphics[width=\textwidth, trim=0 20 0 5, clip]{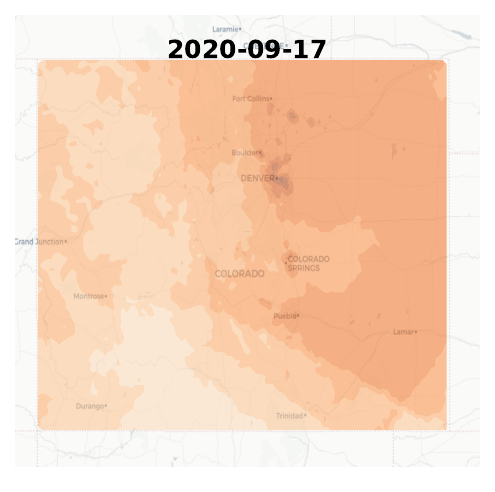}
        \label{fig:cameron_d}
    \end{subfigure}
    \hfill
    \begin{subfigure}[b]{0.30\textwidth}
        \centering
        \caption{}
        \includegraphics[width=\textwidth, trim=0 20 0 5, clip]{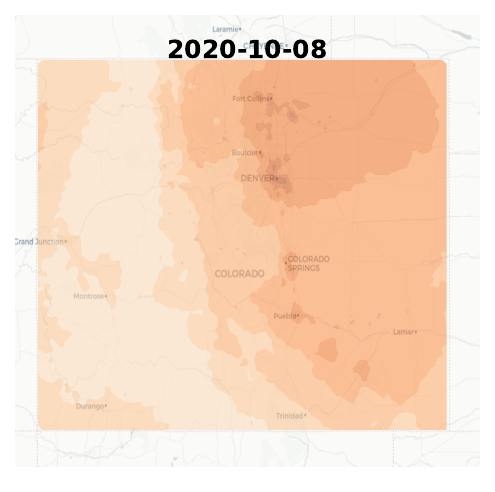}
        \label{fig:cameron_e}
    \end{subfigure}
    \hfill
    \begin{subfigure}[b]{0.3\textwidth}
        \centering
        \caption{}
        \includegraphics[width=\textwidth, trim=0 20 0 5, clip]{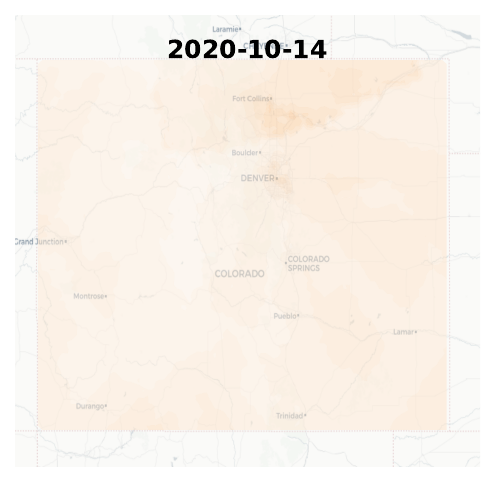}
        \label{fig:cameron_f}
    \end{subfigure}
    \vspace{-0.5em}
    \caption{%
    Cameron Peak Fire PM2.5 evolution. 
    (a)~August 13, 2020: Day of ignition.
    (b)~August 23, 2020: The fire has grown to around 23,000 acres and is about 5\% contained.
    (c)~September 6, 2020: Labor Day weekend blowup (adds roughly 70,000 acres from September 6 to 8).
    (d)~September 17, 2020: Sustained activity as the fire exceeds 125,000 acres.
    (e)~October 8, 2020: Red flag conditions with high wind.
    (f)~October 14, 2020: Peak expansion---becomes the largest wildfire in Colorado history with over 23,000 acres burned in a single day.}
    \label{fig:cameron_peak_fire}
\end{figure}

The Cameron Peak Fire, which ignited on August 13, 2020, became the largest wildfire in Colorado's recorded history and burned over 200,000 acres before it was fully contained in early December 2020.
To evaluate our model's ability to capture wildfire-driven air quality dynamics, we analyzed predictions during periods of intense fire activity.
Figure~\ref{fig:cameron_peak_fire} shows the Senseiver PM2.5 predictions of the spatiotemporal progression of pollution associated with the fire.

The model captures distinct smoke plumes emanating from the fire's location in north-central Colorado, exhibiting coherent spatial gradients consistent with atmospheric transport.
The predictions show high concentrations in mountain valleys, which reflects topographic channeling effects, as well as the day-to-day variability driven by changing meteorological conditions and fire behavior.

Comparison with satellite imagery from Colorado State University's Cooperative Institute for Research in the Atmosphere satellite library\footnote{\url{https://satlib.cira.colostate.edu/event/cameron-peak-fire-2/}} reveals a strong correspondence between our predicted PM2.5 distributions and visible smoke plumes.
The spatial alignment between the Senseiver model's predictions and satellite-observed smoke transport validates the model's ability to capture fire-driven air quality dynamics. Ground-truth sensor measurements further confirm the model's accuracy in predicting localized PM2.5 concentrations during this extreme fire event.

\bmhead{Acknowledgments}

\thanks{This material is based upon work supported by the US Department of Energy (DOE), Office of Science, Office of Advanced Scientific Computing under award no. DE-SC-ERKJ422.
This manuscript has been authored by UT-Battelle LLC under contract DE-AC05-00OR22725 with the DOE. The publisher acknowledges the US government license to provide public access under the DOE Public Access Plan (http://energy.gov/downloads/doe-public-access-plan).}

\bibliography{sn-bibliography}% 

\end{document}